\documentclass[]{spie}  
 
\usepackage{amsmath,amsfonts,amssymb}
\usepackage{graphicx}
\usepackage[colorlinks=true, allcolors=blue]{hyperref}
\usepackage{makecell}
\usepackage{multirow}
\usepackage{subfig}
\usepackage[T1]{fontenc}
\usepackage[utf8]{inputenc}
\usepackage[english, norsk]{babel}
\usepackage{lmodern}

\title{Towards a Video Quality Assessment based Framework for Enhancement of Laparoscopic Videos}

\author[a]{Zohaib Amjad Khan}
\author[a]{Azeddine Beghdadi}
\author[b]{Faouzi Alaya Cheikh}
\author[a]{Mounir Kaaniche}
\author[c,d]{Egidijus Pelanis}
\author[b]{Rafael Palomar}
\author[c,d,e]{{\AA}smund Avdem Fretland}
\author[c,d,e]{Bj{\o}rn Edwin}
\author[c,f]{Ole Jakob Elle}

\affil[a]{L2TI-Institut Galil\'ee, Universit\'e Paris 13, Villetaneuse, France}
\affil[b]{Norwegian Colour and Visual Computing Lab, NTNU, Gj{\o}vik, Norway}
\affil[c]{The Intervention Centre, Oslo University Hospital – Rikshospitalet, Oslo, Norway}
\affil[d]{Institute of Clinical Medicine, University of Oslo, Oslo, Norway}
\affil[e]{Department of HPB Surgery, Oslo University Hospital – Rikshospitalet, Oslo, Norway}
\affil[f]{Department of Informatics, University of Oslo, Oslo, Norway}

\authorinfo{Further author information: (Send correspondence to Z.A. Khan)\\Z.A. Khan: E-mail: zohaibamjad.khan@univ-paris13.fr, \\ A. Beghdadi: E-mail: beghdadi@univ-paris13.fr}

\begin{document} 
\maketitle

\begin{abstract}
Laparoscopic videos can be affected by different distortions which may impact the performance of surgery and introduce surgical errors. In this work, we propose a framework for automatically detecting and identifying such distortions and their severity using video quality assessment. There are three major contributions presented in this work (i) a proposal for a novel video enhancement framework for laparoscopic surgery; (ii) a publicly available database for quality assessment of laparoscopic videos evaluated by expert as well as non-expert observers and (iii) objective video quality assessment of laparoscopic videos including their correlations with expert and non-expert scores.
\end{abstract}


\section{INTRODUCTION}
\label{sec:intro}  

A satisfactory video quality is an important requirement for achieving optimal conditions for laparoscopic surgery. The distortions in a laparoscopic video not only affect a surgeon's visibility but also degrade the results of subsequent computational tasks in robot-assisted surgery and image-guided navigation systems \cite{sanchez2011laparoscopic}. 
Examples of such tasks are segmentation \cite{bodenstedt2018real,voros2007automatic}, 
instrument tracking \cite{bouget2017vision,zhou2014visual}, 
and augmented reality \cite{bernhardt2017status}. 

Laparoscopic videos may be affected by different kinds of distortions during the surgery, resulting in a loss of visual quality. These are mainly a result of technical problems in the equipment \cite{verdaasdonk2007problems} or side-effects of the instruments being used (e.g. smoke with diathermy). To deal with such problems, most of the suggested prevalent solutions rely on making some changes to the technical equipment using one of the many available troubleshooting options as also highlighted in Ref.~\citenum{siddaiah2017technical}. However, all such solutions are time-consuming and may not always solve the problem at hand requiring eventually a specialist technician or a change in apparatus \cite{verdaasdonk2007problems}.

\begin{figure} [t]
	\begin{center}
			\includegraphics[width = 0.95 \textwidth]{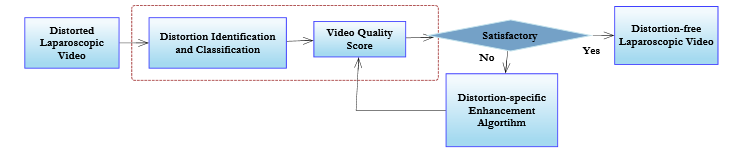}
	\end{center}
	\caption[flow] 
	{ \label{fig_flow} 
		Flowchart of the proposed framework for laparoscopic video enhancement with quality control}
\end{figure} 

In this work, we propose a computational framework for automatic detection and correction of video quality for laparoscopic surgery (Figure \ref{fig_flow}).  The proposed framework consists of a video quality assessment (VQA) module followed by an enhancement module. This work solely focuses on the video quality assessment part (dotted red area in Figure \ref{fig_flow}) which is composed of two stages namely distortion detection/classification and video quality score evaluation. Such hybrid two-step quality assessment techniques are not new and have already been proposed for natural images \cite{chetouani2012hybrid}.

Quality assessment of videos, if done subjectively, is time-consuming and hence not feasible. In order to assess video quality automatically, objective metrics are needed. However, the effectiveness of a designed objective metric can only be evaluated by using a database of videos annotated with subjective scores \cite{winkler2012analysis}. 
Unfortunately, to the best of our knowledge, there is no such database of laparoscopic videos available publicly. Hence, there is currently a big gap to be filled in the field of medical visual quality assessment. This work also aims to fill this gap by proposing a new database which is dedicated to laparoscopic video quality assessment (Available at: \href{https://drive.google.com/file/d/1SoONeacp9vvihTY7zmWssG_cnVzx16oq/view?usp=sharing}{LVQ Database}) \footnote[1]{URL: \url{https://drive.google.com/file/d/1SoONeacp9vvihTY7zmWssG_cnVzx16oq/view?usp=sharing}.}

\section{Laparoscopic Video Quality (LVQ) Database}

Our database called the Laparoscopic Video Quality (LVQ) database consists of a total of 10 reference videos, each of 10 seconds. Each reference video is distorted by five different kinds of distortions with four different levels, resulting in a total of 200 videos. The resolution of the videos is 512 $\times$ 288 with a 16:9 aspect ratio and a frame-rate of 25 fps. Moreover, we have used uncompressed avi format for the videos so as to avoid any kind of unwanted compression artefacts like blocking. In the following sections, we describe in more details the construction of our database.

\subsection{Selection of Videos}
\label{sec:video_sel}

\begin{figure}[tpb]
	\centering
	\subfloat[Bleeding (BL).\label{fig:bleed}]{\includegraphics[width=0.17\textwidth]{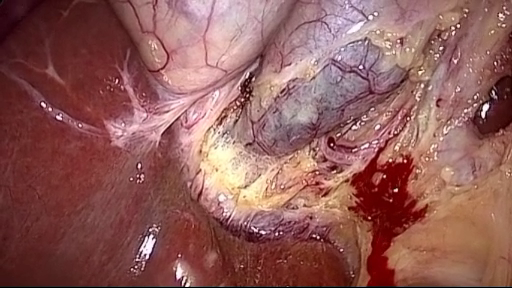}}\hspace{0.01\textwidth}
	\subfloat[Grasp and Burn (GB).\label{fig:grasp_burn}] {\includegraphics[width=0.17\textwidth]{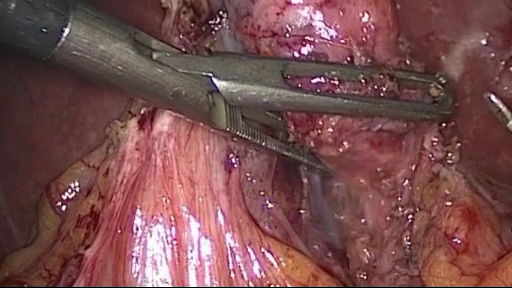}}\hspace{0.01\textwidth}
	\subfloat[Multiple instruments (MI).\label{fig:multi_instruments}]{\includegraphics[width=0.17\textwidth]{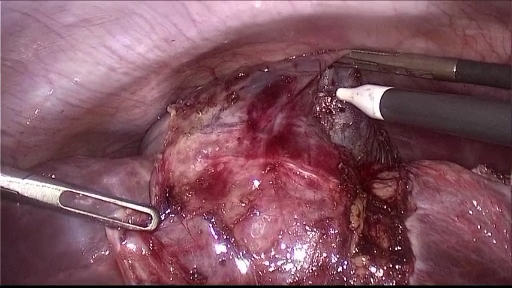}}\hspace{0.01\textwidth}
	\subfloat[Irrigation (IR).\label{fig:irrigate}]{\includegraphics[width=0.17\textwidth]{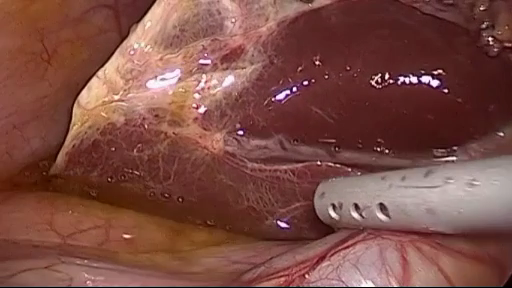}}\hspace{0.01\textwidth}
	\subfloat[Clipping (CL).\label{fig:clip}] {\includegraphics[width=0.17\textwidth]{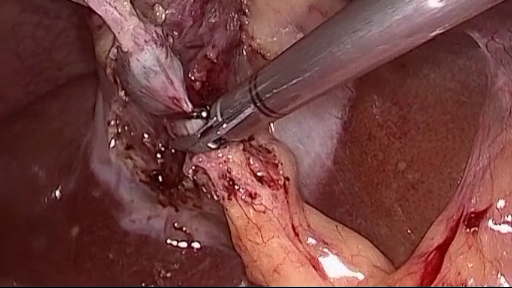}}\hspace{0.01\textwidth}
	\subfloat[Stretching away (SA).\label{fig:stretch_away}]{\includegraphics[width=0.17\textwidth]{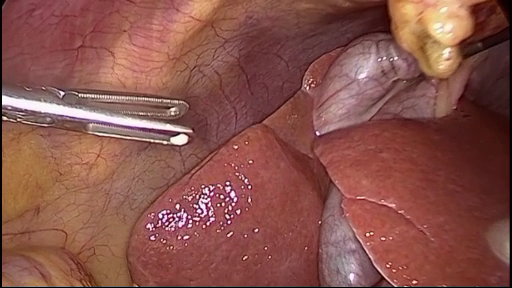}}\hspace{0.01\textwidth}
	\subfloat[Cutting (CU).\label{fig:cut}]{\includegraphics[width=0.17\textwidth]{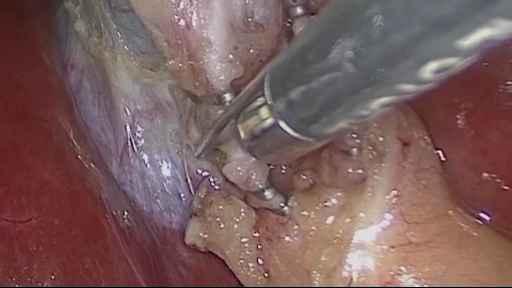}}\hspace{0.01\textwidth}
	\subfloat[Grasping and stretching forward (SF).\label{fig:stretch_forward}] {\includegraphics[width=0.17\textwidth]{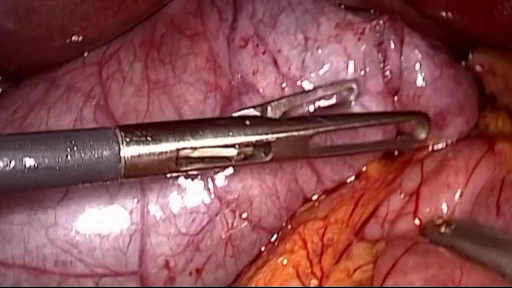}}\hspace{0.01\textwidth}
	\subfloat[Organ Extraction (OE).\label{fig:collect}]{\includegraphics[width=0.17\textwidth]{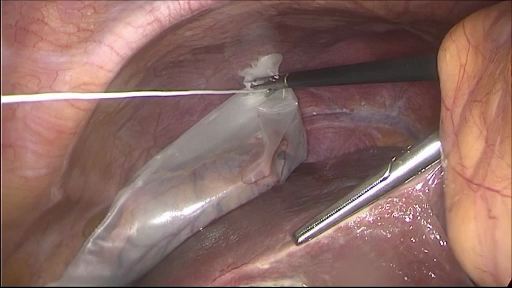}}\hspace{0.01\textwidth}
	\subfloat[Burning (BU).\label{fig:burn}]{\includegraphics[width=0.17\textwidth]{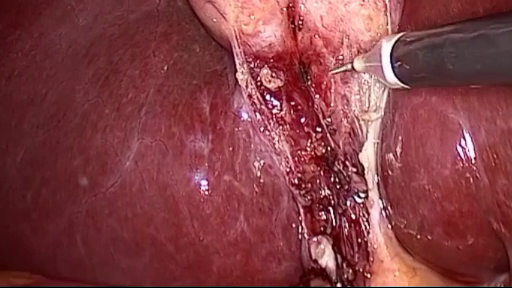}}\hspace{0.01\textwidth}
	\caption{\label{fig:ref_videos} One frame from each of the reference videos in the LVQ database.} 
\end{figure}

For the database, ten different videos of laparoscopic cholecystectomy are selected as reference. These videos are extracted from Cholec80 dataset \cite{twinanda2017endonet} and are shown in Figure~\ref{fig:ref_videos}. The selection of the videos is made with an attempt to include maximum possible variations of scene content and temporal information. For scene content, ten different categories are chosen as illustrated in Figure~\ref{fig:ref_videos}. These are bleeding (BL), grasping and burning (GB), multiple instruments (MI), irrigation (IR), clipping (CL), stretching away (SA), cutting (CU), stretching forward (SF), organ extraction (OE) and burning (BU).

\subsection{Creation of Distorted Videos}
We have chosen five different distortions for our database. These distortions, which are among the most common affecting a laparoscopic video, are the smoke, noise, uneven illumination, blur due to defocus and blur due to motion. In order to simulate each of these distortions, we have applied appropriate mathematical models to every frame of the reference video. For generating four levels of severity for each distortion, we have modified the relevant parameters of these models. For this initial work, we have not considered time-variatons in distortions. Hence, in each video there is one single distortion at same level throughout. 

We have used MATLAB to simulate all the distortions with different levels for our database. For defocus blur, a symmetric low-pass Gaussian filter was applied to each video frame. The filter size and the standard deviation were changed to generate different levels of this distortion. For motion blur, MATLAB motion filter was used with variations in filter length parameter to generate different severity levels. Similarly, built-in MATLAB function for noise was used to add additive white Gaussian noise. In this case, variance of the Gaussian model was adjusted to generate multiple levels of noisy videos.

For uneven illumination, a special grayscale mask was generated having a circular bright region of high intensity values. The areas surrounding the circular region were generated with decreasing intensities which were attenuated as a function of distance from the center of the bright region. The multiplication of this mask with the original frame gave us the unevenly illuminated distorted frame. By changing the two parameters of the center location of the bright region and its area, we generated four different levels for unveven illumination.

In order to generate smoke, we have used a well-known method of video editing called the screen blending. In this technique, a smoke-only video having a black background is combined with the reference video in such a way that black areas produce no change to the original video while the brighter areas overlay the original ones. Using different opacity levels for the smoke video we have generated four different levels of severity for smoke.

\subsection{Subjective Tests}

For the subjective testing, we have used pairwise-comparison protocol\cite{itu2008p910,qureshi2017towards}. For each observer, we randomly displayed all possible pair combinations of distorted videos, such that each pair had two videos from the same category and the same distortion type, with only difference being the severity level. This corresponded to 6 pair-wise comparisons per reference video for each distortion type. 

The observers had the choice to give an equal score to the videos if they perceived so. For each comparison, the preferred video was given one point. In case of an equal choice, a score of 0.5 was given to each displayed video. The observers were shown each video once and they had the choice to see the video again if required. In the end, the scores for each video from all the observers were added. Finally the Mean Opinion Score (MOS) for the $i$-th video was obtained by averaging the total score for that video over number of observers $N$ \cite{ponomarenko2015image}.
\begin{equation}
\label{eq:mos}
MOS_i = \frac{1}{N} \sum_{j=1}^N score_{ij}
\end{equation}

\begin{figure} [t]
	\begin{center}
		\begin{tabular}{c} 
			\includegraphics[width = 0.5 \textwidth]{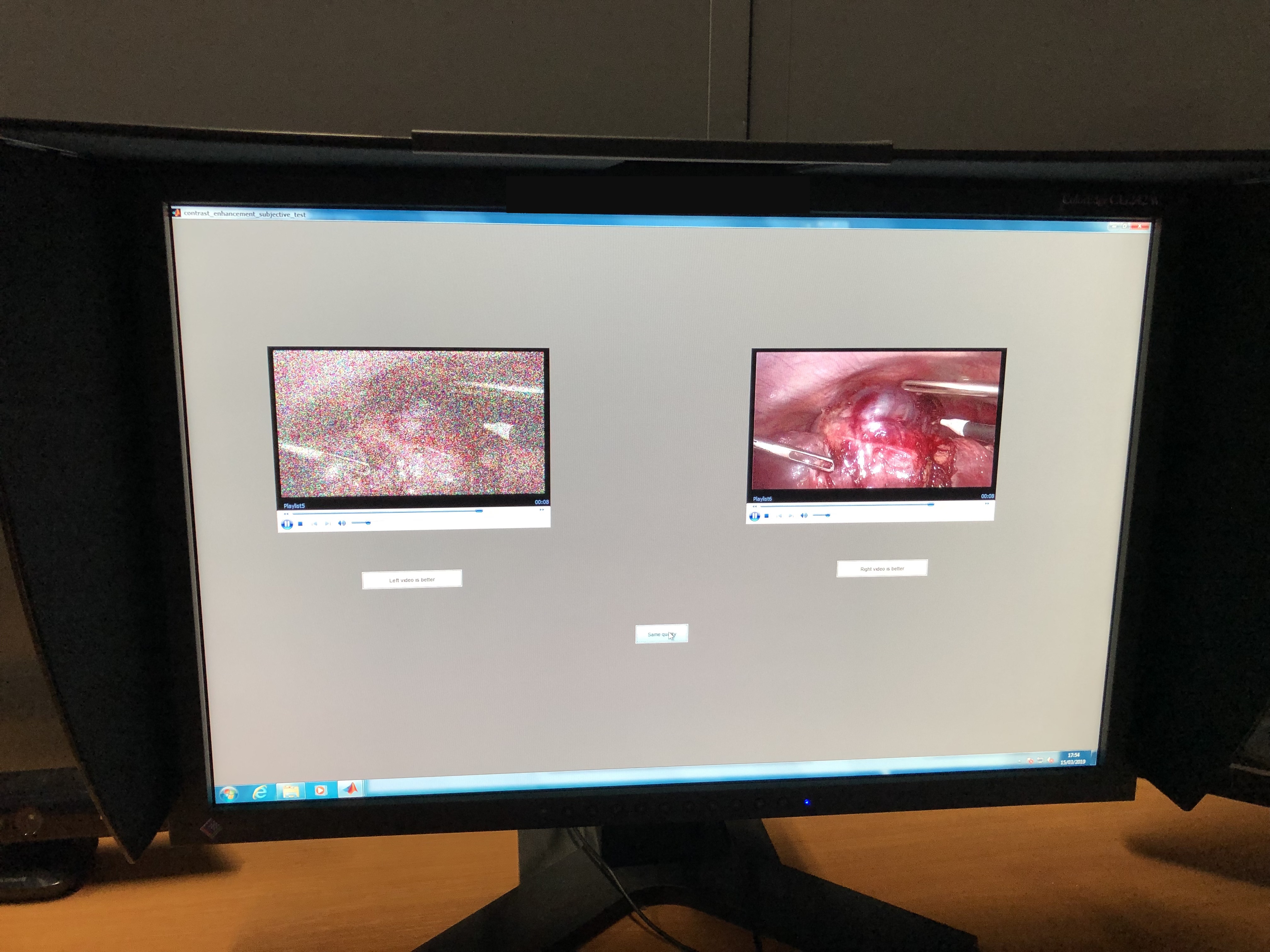}
		\end{tabular}
	\end{center}
	\caption[pc] 
	{ \label{fig_pc} 
		Setup for subjective tests}
\end{figure}

In order to perform the subjective tests, a calibrated 24.1 inch LCD monitor was used. The observers were forced to perform the experiments at a fixed distance of twice the screen height which is equivalent in our case to be 4.5 times the image height of the image. The setup for the subjective tests is shown in Figure~\ref{fig_pc}. All the observers had either normal vision or corrected to normal vision and they were undergone a pre-screening procedure for color vision and visual acuity.

\section{Laparoscopic Video Quality Assessment}

In order to apply the pertinent enhancement method with suitable parameters in our proposed video enhancement framework, VQA module should be able to detect the type of distortion affecting a video as well as its severity. For this reason, our VQA consists of a distortion identification step followed by the quality score estimation.

\subsection{Distortion Classification}
As a first step, for each kind of distortion, we have chosen a distortion-specific classification method. In our selection of these methods, we have gone for no-reference, opinion-unaware, accurate and computationally less expensive methods to allow for real-time performance. Below are the details of the classification methods used for each kind of distortion.

\subsubsection{Motion and defocus blur}
For the two kinds of blur, we have used Perceptual Blur Index (PBI)\cite{chetouani2009new} with threshold as the classifier. PBI is a quality metric for estimating the level of blur in an image. It is based on the way Human Visual System (HVS) perceives addition of blur to an already blurred image and to a sharp one differently . The perceptual difference is more pronounced for the latter case. It is defined in terms of the difference between total radial energy of the input image $RE(w)$ and that of its binomial filtered version $RE_f(w)$ as 
\begin{equation}
\label{eq:pbi}
PBI = \log (\frac{1}{w_{max}} \sum_w |RE(w) - RE_f(w)|)
\end{equation}
where $w_{max}$ is the maximal frequency.

\subsubsection{Smoke}In order to detect if there is smoke in a video, we have used Saturation Analysis (SAN) classifier \cite{leibetseder2017real}. SAN classifier uses the histogram of saturation channel of a frame to detect smoke. If the majority of bin values in histogram $hist$ are below the chosen threshold $t_c$, the video frame is classified to have smoke in it. The threshold used is $t_c=0.35$ as suggested in the original work \cite{leibetseder2017real}. The probability of an image having smoke $p_{(S)}$ and no smoke $p_{(NS)}$ are therefore defined as
\begin{equation}
\label{eq:san}
p_{(S)} = \frac{1}{|hist|} \sum_{\substack{i=0\\b\in hist\\ t\leq t_c}} b_i
\end{equation}
\begin{equation}
\label{eq:san2}
p_{(NS)} = 1 - p_{(S)}
\end{equation}
where $b_i$ is the $i$-th bin value of the histogram $hist$.

\subsubsection{Noise}For noise classification we have chosen the fast noise variance estimator \cite{immerkaer1996fast} with threshold. In this method, the standard deviation of additive white Gaussian noise in an image is estimated using a noise estimation mask. This suggested mask, $M_N$ has been generated using a difference of two $3\times3$ masks, each approximating the Laplacian of an image. For an image $I$ with width $W$ and height $H$, the estimated standard deviation $\sigma_n$ of noise is estimated as\cite{immerkaer1996fast}:
\begin{equation}
\label{eq:san}
\sigma_n = \sqrt{\frac{\pi}{2}}\frac{1}{6(W-2)(H-2)} \sum_{x,y} |I(x,y)*M_N|
\end{equation}

\subsubsection{Uneven illumination}In order to detect whether a video is affected by uneven illumination or not, we have developed a novel classifier which makes use of statistics of the luminance component of an image. For an unevenly illuminated laparoscopic video frame, there are some dark regions in the image which tend to increase the range of values for the luminance component in an image, while reducing the mean luminance value of the image at the same time. Making use of these trends, we have proposed a new classifier that uses a threshold on the Luminance Mean to Range (LMR) ratio, defined simply as the ratio of mean luminance value to that of the range of luminance values in an image. For an image with $N_p$ pixels and with luminance component $Y$, this index can be defined as
\begin{equation}
\label{eq:uneven}
LMR = \frac{\frac{1}{N_p} \sum_{i=0}^{N_p} Y_i}{max(Y) - min(Y)}
\end{equation}
An image with a $LMR$ value smaller than a pre-defined threshold can be classified to have been affected by uneven illumination.  

\subsection{Video Quality Score}
Once the distortion is identified in a video, we can evaluate its severity using a quality score. To this effect, we have selected 3 different metrics which are often used as benchmarks for natural images and videos. These are PSNR, SSIM\cite{wang2004image} and VIF\cite{sheikh2006image}. However, for laparoscopic videos, there is usually no ground truth available and a no-reference (NR) metric makes more sense. For this reason, we have also included NR metrics. However, due to the limited number of good NR metrics for videos, we have only selected one of the more recent NR metrics dedicated to opinion-unaware VQA called VIIDEO\cite{mittal2015completely}. We have also included two NR image quality metrics BRISQUE\cite{mittal2012no} and NIQE\cite{mittal2012making}. For both of these, we have used the mean metric value from all frames as the score for the video.

\begin{figure} [htbp]
	\begin{center}
		\begin{tabular}{c} 
			\includegraphics[width = 0.5 \textwidth]{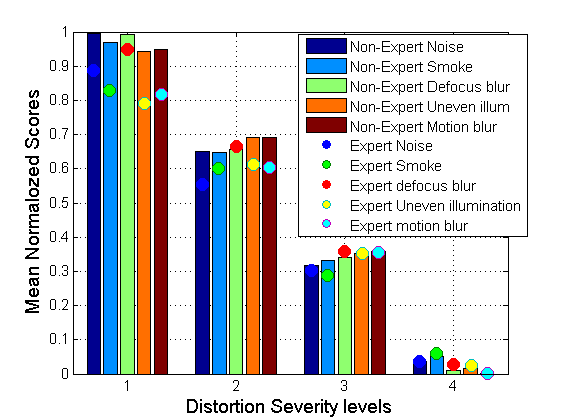}
		\end{tabular}
	\end{center}
	\caption[pc] 
	{ \label{figMOS_compare} 
		Comparison of subjective scores for experts and non-experts}
\end{figure} 

 \begin{table}[ht]
	\caption{PLCC for \textbf{non-expert} scores in LVQ Database (best two values in bold for each column)} 
	\label{pearson_nonexpert}
	\begin{center}       
		\begin{tabular}{|l|l|l|l|l|l|l|}
			\hline
			\rule[-1ex]{0pt}{2.5ex} \textbf{Metric}&\textbf{Noise}&\makecell{\textbf{Defocus}\\\textbf{Blur}} &\makecell{\textbf{Motion}\\\textbf{Blur}}&\makecell{\textbf{Uneven}\\ \textbf{illumination}}&\textbf{Smoke}&\textbf{Overall}\\ 
			\hline			
			\rule[-1ex]{0pt}{3.5ex} 
			\textbf{PSNR}&\textbf{0.9968}&0.8166&0.8199&0.9561&\textbf{0.9811}&0.6054\\	
			\hline			
			\rule[-1ex]{0pt}{3.5ex} \textbf{SSIM}&0.9690&0.7388 &\textbf{0.8861}&\textbf{0.9926}&0.9165&\textbf{0.6123} \\	
			\hline					
			\rule[-1ex]{0pt}{3.5ex} \textbf{VIF}&\textbf{0.9925}&\textbf{0.9764}&\textbf{0.9713}&\textbf{0.9919}&\textbf{0.9853}&\textbf{0.6267} \\	
			\hline
			\rule[-1ex]{0pt}{3.5ex} \textbf{BRISQUE}& 0.9803   &   0.9646  &  0.4090  &  0.3142  &  0.3735 & 0.4593
			\\	
			\hline	
			\rule[-1ex]{0pt}{3.5ex} \textbf{NIQE}&   0.9783  &    \textbf{0.9880} &      0.7704    &  0.6618   &   0.3238 & 0.4242 \\	
			\hline
			\rule[-1ex]{0pt}{3.5ex} \textbf{VIIDEO}&0.8749&0.3549&0.4998&0.3983&0.4214&0.3842
			\\	
			\hline	
		\end{tabular}
	\end{center}
\end{table}

 \begin{table}[ht]
	\caption{SROCC for \textbf{non-expert} scores in LVQ Database (best two values in bold for each column)} 
	\label{spearman_nonexpert}
	\begin{center}       
		\begin{tabular}{|l|l|l|l|l|l|l|}
			\hline
			\rule[-1ex]{0pt}{2.5ex} \textbf{Metric}&\textbf{Noise}&\makecell{\textbf{Defocus}\\\textbf{Blur}} &\makecell{\textbf{Motion}\\\textbf{Blur}}&\makecell{\textbf{Uneven}\\ \textbf{illumination}}&\textbf{Smoke}&\textbf{Overall}\\ 
			\hline			
			\rule[-1ex]{0pt}{3.5ex} 
			\textbf{PSNR}&0.9594&0.7773&0.8163&0.9372&\textbf{0.9439}&0.5775 \\	
			\hline			
			\rule[-1ex]{0pt}{3.5ex} \textbf{SSIM}&0.9509 &0.7157&\textbf{0.8941}&\textbf{0.9502}&0.8987&\textbf{0.5914} \\	
			\hline					
			\rule[-1ex]{0pt}{3.5ex} \textbf{VIF}&\textbf{0.9636}&\textbf{0.9417}&\textbf{0.9433}&\textbf{0.9391}&\textbf{0.9316}&\textbf{0.6228} \\	
			\hline
			\rule[-1ex]{0pt}{3.5ex} \textbf{BRISQUE}  &  0.9571 &    0.9332  &  0.3564  & 0.2980  &  0.4041 & 0.4304
			\\	
			\hline	
			\rule[-1ex]{0pt}{3.5ex} \textbf{NIQE}  &  \textbf{0.9640}   &   \textbf{0.9514}  &  0.6101    &  0.5416  &   0.3589 & 0.3731\\	
			\hline
			\rule[-1ex]{0pt}{3.5ex} \textbf{VIIDEO}&0.8600&0.3138&	0.379&0.3888&0.3866&0.3416
			\\	
			\hline	
		\end{tabular}
	\end{center}
\end{table}

\begin{table}[htbp]
	\caption{PLCC for \textbf{expert} scores in LVQ Database (best two values in bold for each column)}
	\label{pearson_expert}
	\begin{center}
		\begin{tabular}{|l|l|l|l|l|l|l|}
				\hline
				\rule[-1ex]{0pt}{3.5ex} \textbf{Metric}&\textbf{Noise}&\makecell{\textbf{Defocus}\\\textbf{Blur}} &\makecell{\textbf{Motion}\\\textbf{Blur}}&\makecell{\textbf{Uneven}\\ \textbf{illumination}}&\textbf{Smoke}&\textbf{Overall}\\ 
				\hline					
				\rule[-1ex]{0pt}{3.5ex} \textbf{PSNR}&\textbf{0.9939}&0.8146&0.8226&0.9452&\textbf{0.9777}&\textbf{0.6853} \\	
				\hline					
				\rule[-1ex]{0pt}{3.5ex} \textbf{SSIM}&0.9706&0.7358&\textbf{0.8827}&\textbf{0.9847}&0.9116&0.5732 \\	
				\hline					
				\rule[-1ex]{0pt}{3.5ex} \textbf{VIF}&\textbf{0.9896}&\textbf{0.9806}&\textbf{0.9708}&\textbf{0.9878}&\textbf{0.9808}&\textbf{0.5909}  \\	
				\hline	
				\rule[-1ex]{0pt}{3.5ex} \textbf{BRISQUE}& 0.9761  &  0.9623  &  0.4208  &   0.2973 &   0.4009  & 0.4434 \\	
				\hline	
				\rule[-1ex]{0pt}{3.5ex} \textbf{NIQE}&  0.9741  &   \textbf{0.9883}   &  0.7836    &  0.6655   & 0.4301 & 0.4407  \\	
				\hline	
				\rule[-1ex]{0pt}{3.5ex} \textbf{VIIDEO}&0.8658&0.3498&0.5136&0.4035&0.4195&0.3744
				\\	
				\hline
		\end{tabular}	
	\end{center}
\end{table}

\begin{table}[htbp]
	\caption{SROCC for \textbf{expert} scores in LVQ Database (best two values in bold for each column)}
	\label{spearman_expert}
	\begin{center}
		\begin{tabular}{|l|l|l|l|l|l|l|}
			\hline
			\rule[-1ex]{0pt}{3.5ex} \textbf{Metric}&\textbf{Noise}&\makecell{\textbf{Defocus}\\\textbf{Blur}} &\makecell{\textbf{Motion}\\\textbf{Blur}}&\makecell{\textbf{Uneven}\\ \textbf{illumination}}&\textbf{Smoke}&\textbf{Overall}\\ 
			\hline					
			\rule[-1ex]{0pt}{3.5ex} \textbf{PSNR}&0.9579&0.7836&0.7977&0.9530&\textbf{0.9478}&\textbf{0.6914} \\	
			\hline					
			\rule[-1ex]{0pt}{3.5ex} \textbf{SSIM}&0.9435&0.7320&\textbf{0.8802}&\textbf{0.9580}&0.8817&\textbf{0.5653} \\	
			\hline					
			\rule[-1ex]{0pt}{3.5ex} \textbf{VIF}&\textbf{0.9592}&\textbf{0.9555}&\textbf{0.9376}&\textbf{0.9534} &\textbf{0.9459}&0.5642 \\	
			\hline	
			\rule[-1ex]{0pt}{3.5ex} \textbf{BRISQUE} & 0.9527  & 0.9355   &   0.3994  &   0.2634  &  0.4355 & 0.3842\\	
			\hline	
			\rule[-1ex]{0pt}{3.5ex} \textbf{NIQE} &   \textbf{0.9594}  &   \textbf{0.9443}   &  0.7028    &  0.5605   &  0.3382 & 0.3674\\	
			\hline	
			\rule[-1ex]{0pt}{3.5ex} \textbf{VIIDEO}&0.8822&0.3023&	0.3915&0.4281&0.4416&0.3334
			\\	
			\hline
		\end{tabular}	
	\end{center}
\end{table}

\section{Results and Discussion}

In total, thirty non-expert and ten expert observers performed the subjective tests for the database. Both the expert and non-expert observers were considered as two separate groups. For each group, outliers were first detected based on non-transitivity. This corresponded to one subject in each group. The preference matrices for remaining subjects in the two groups were then compiled and aggregated to obtain subjective scores.

Figure ~\ref{figMOS_compare} shows a comparison between expert and non-expert mean normalized scores for LVQ database. From the figure, we can clearly see how experts perceive quality much differently for all distortions except for defocus blur. The difference is more pronounced for less distorted videos (levels 1 and 2) suggesting how even the slightest level of distortion affects the perception of a video for experts (who are more task-oriented).

To evaluate the performance of our selected classification methods, all the videos in our database were passed through these classifiers. The results obtained for classification accuracies were: smoke - 87\%; motion blur - 89\%; defocus blur - 91.5\%; noise - 100\% and uneven illumination classifiers - 88.5\%.  

Furthermore, in order to assess whether an existing objective video quality metric correlates well or not with subjective scores, Pearson Linear Correlation Coefficient (PLCC) and Spearman Rank Order Correlation Coefficient (SROCC) were evaluated for the metric scores after performing a non-linear regression with a 5-parameter logistic function. From Tables~\ref{pearson_nonexpert} and \ref{pearson_expert}, we can see that none of the objective metrics perform well when overall correlations are evaluated, with maximum PLCC value of 0.6267 with VIF for non-experts and a value of 0.6853 with PSNR for experts. 

However, with individual distortion types, VIF correlates much better with subjective scores as compared to others for both groups and for all the distortions. Among the NR metrics, both NIQE and BRISQUE give good results for noise and defocus blur, with NIQE being better of the two for motion blur and uneven illumination. However, VQA specific method VIIDEO performs poorly for all distortions except for the noise. 

All these results are very significant as they imply that none of these metrics are generic or non-distortion specific for the kind of videos and distortions encountered in medical domain.
Moreover, these results also show a difference with respect to their correlation with expert and non-expert scores. To be more specific, if we compare the results of experts and non-experts, we can see that generally all the metrics tend to correlate better with non-expert opinion as compared to expert opinion.

\section{Conclusion}

In this work, we have proposed a novel computational framework for laparoscopic video enhancement based on video quality assessment. Especially, we have taken a major initiative for quality assessment of laparoscopic videos by creating a database with subjective quality scores not only from normal observers but also from medical experts. Our initial results show that the existing NR metrics for video quality assessment are not sufficient especially in context of laparoscopic videos. Moreover, we have observed that experts and non-experts differ in their opinions on video quality assessment and new no-reference metrics are required to model expert opinion. In this regards, the constructed LVQ database is an important step to address the above challenges in a future work and in facilitating development of new VQA metrics in the medical imaging context.

\acknowledgments 
 
This research work is part of a project that has received funding from the European Union's Horizon 2020 research and innovation programme under grant agreement No 722068. We would also like to thank all the observers who took part in the subjective experiments especially all the medical experts from Oslo University Hospital (Rikshospitalet), Norway.

\bibliography{report} 

\begin{thebibliography}{10}

\bibitem{sanchez2011laparoscopic}
Sanchez-Gonzalez, P., Cano, A.~M., Oropesa, I., Sanchez-Margallo, F.~M., Pozo,
  F.~D., Lamata, P., and G{\'o}mez, E.~J., ``Laparoscopic video analysis for
  training and image-guided surgery,'' {\em Minimally Invasive Therapy \&
  Allied Technologies}~{\bf 20}(6),  311--320 (2011).

\bibitem{bodenstedt2018real}
Bodenstedt, S., Ohnemus, A., Katic, D., Wekerle, A.-L., Wagner, M., Kenngott,
  H., M{\"u}ller-Stich, B., Dillmann, R., and Speidel, S., ``Real-time
  image-based instrument classification for laparoscopic surgery,'' {\em arXiv
  preprint arXiv:1808.00178}  (2018).

\bibitem{voros2007automatic}
Voros, S., Long, J.-A., and Cinquin, P., ``Automatic detection of instruments
  in laparoscopic images: A first step towards high-level command of robotic
  endoscopic holders,'' {\em The International Journal of Robotics
  Research}~{\bf 26}(11-12),  1173--1190 (2007).

\bibitem{bouget2017vision}
Bouget, D., Allan, M., Stoyanov, D., and Jannin, P., ``Vision-based and
  marker-less surgical tool detection and tracking: a review of the
  literature,'' {\em Medical image analysis}~{\bf 35},  633--654 (2017).

\bibitem{zhou2014visual}
Zhou, J. and Payandeh, S., ``Visual tracking of laparoscopic instruments,''
  {\em Journal of Automation and Control Engineering Vol}~{\bf 2}(3) (2014).

\bibitem{bernhardt2017status}
Bernhardt, S., Nicolau, S.~A., Soler, L., and Doignon, C., ``The status of
  augmented reality in laparoscopic surgery as of 2016,'' {\em Medical image
  analysis}~{\bf 37},  66--90 (2017).

\bibitem{verdaasdonk2007problems}
Verdaasdonk, E.~G., Stassen, L.~P., van~der Elst, M., Karsten, T.~M., and
  Dankelman, J., ``Problems with technical equipment during laparoscopic
  surgery,'' {\em Surgical endoscopy}~{\bf 21}(2),  275--279 (2007).

\bibitem{siddaiah2017technical}
Siddaiah-Subramanya, M., Nyandowe, M., and Tiang, K.~W., ``Technical problems
  during laparoscopy: a systematic method of troubleshooting for surgeons,''
  {\em Innovative Surgical Sciences}~{\bf 2}(4),  233--237 (2017).

\bibitem{chetouani2012hybrid}
Chetouani, A., Beghdadi, A., and Deriche, M., ``A hybrid system for distortion
  classification and image quality evaluation,'' {\em Signal Processing: Image
  Communication}~{\bf 27}(9),  948--960 (2012).

\bibitem{winkler2012analysis}
Winkler, S., ``Analysis of public image and video databases for quality
  assessment,'' {\em IEEE Journal of Selected Topics in Signal Processing}~{\bf
  6}(6),  616--625 (2012).

\bibitem{twinanda2017endonet}
Twinanda, A.~P., Shehata, S., Mutter, D., Marescaux, J., De~Mathelin, M., and
  Padoy, N., ``Endonet: a deep architecture for recognition tasks on
  laparoscopic videos,'' {\em IEEE transactions on medical imaging}~{\bf
  36}(1),  86--97 (2017).

\bibitem{itu2008p910}
ITU-T, R., ``P910,'' {\em Subjective video quality assessment methods for
  multimedia applications}  (2008).

\bibitem{qureshi2017towards}
Qureshi, M.~A., Beghdadi, A., and Deriche, M., ``Towards the design of a
  consistent image contrast enhancement evaluation measure,'' {\em Signal
  Processing: Image Communication}~{\bf 58},  212--227 (2017).

\bibitem{ponomarenko2015image}
Ponomarenko, N., Jin, L., Ieremeiev, O., Lukin, V., Egiazarian, K., Astola, J.,
  Vozel, B., Chehdi, K., Carli, M., Battisti, F., et~al., ``Image database
  {TID}2013: Peculiarities, results and perspectives,'' {\em Signal Processing:
  Image Communication}~{\bf 30},  57--77 (2015).

\bibitem{chetouani2009new}
Chetouani, A., Beghdadi, A., and Deriche, M., ``A new reference-free image
  quality index for blur estimation in the frequency domain,'' in [{\em 2009
  IEEE International Symposium on Signal Processing and Information Technology
  (ISSPIT)}{\nolinebreak\hspace{0.1em}]},   155--159, IEEE (2009).

\bibitem{leibetseder2017real}
Leibetseder, A., Primus, M.~J., Petscharnig, S., and Schoeffmann, K.,
  ``Real-time image-based smoke detection in endoscopic videos,'' in [{\em
  Proceedings of the on Thematic Workshops of ACM
  Multimedia}{\nolinebreak\hspace{0.1em}]},   296--304 (2017).

\bibitem{immerkaer1996fast}
Immerkaer, J., ``Fast noise variance estimation,'' {\em Computer vision and
  image understanding}~{\bf 64}(2),  300--302 (1996).

\bibitem{wang2004image}
Wang, Z., Bovik, A.~C., Sheikh, H.~R., Simoncelli, E.~P., et~al., ``Image
  quality assessment: from error visibility to structural similarity,'' {\em
  IEEE transactions on image processing}~{\bf 13}(4),  600--612 (2004).

\bibitem{sheikh2006image}
Sheikh, H.~R. and Bovik, A.~C., ``Image information and visual quality,'' {\em
  IEEE Transactions on image processing}~{\bf 15}(2),  430--444 (2006).

\bibitem{mittal2015completely}
Mittal, A., Saad, M.~A., and Bovik, A.~C., ``A completely blind video integrity
  oracle,'' {\em IEEE Transactions on Image Processing}~{\bf 25}(1),  289--300
  (2015).

\bibitem{mittal2012no}
Mittal, A., Moorthy, A.~K., and Bovik, A.~C., ``No-reference image quality
  assessment in the spatial domain,'' {\em IEEE Transactions on image
  processing}~{\bf 21}(12),  4695--4708 (2012).

\bibitem{mittal2012making}
Mittal, A., Soundararajan, R., and Bovik, A.~C., ``Making a “completely
  blind” image quality analyzer,'' {\em IEEE Signal Processing Letters}~{\bf
  20}(3),  209--212 (2012).

\end{thebibliography}
\bibliographystyle{spiebib} 

\end{document}